# IFTT-PIN: Demonstrating the Self-Calibration Paradigm on a PIN-Entry Task

We demonstrate 'IF This Then PIN', a PIN-entry method whose code-pad configuration only pre-exists in the user's mind and is learned on-the-fly by the interface.

Dr. Jonathan Grizou

School of computing, University of Glasgow

Center for Research and Insterdicisplinarity (CRI). Université de Paris

We demonstrate IFTT-PIN, a self-calibrating version of the PIN-entry method introduced in Roth et al. (2004) [1]. In [1], digits are split into two sets and assigned a color respectively. To communicate their digit, users press the button with the same color that is assigned to their digit, which can be identified by elimination after a few iterations. IFTT-PIN uses the same principle but does not pre-assign colors to each button. Instead, users are free to choose which button to use for each color. IFTT-PIN infers both the user's PIN and their preferred button-to-color mapping at the same time, a process called self-calibration. Different versions of IFTT-PIN can be tested at https://jgrizou.github.io/IFTT-PIN/ and a video introduction at https://youtu.be/5I1ibPJdLHM.

# 1 INTRODUCTION

When giving instructions to a machine, humans are funnelled by the interfaces we present to them. Buttons are to be pressed and the letters printed on them indicate what they do. Such interfaces act as translators from user's actions to the symbolic meanings that the machine works with. For example, a remote control transforms the human action of pressing the button marked with a 1 into a digital signal that instructs the machine to display the channel 1 on the screen.

In some cases, the action-to-meaning mapping cannot be pre-defined or needs to be configured by the user. Such personalization requires an explicit calibration procedure to learn an action-to-meaning mapping for each user. This calibration procedure is required due to the chicken-and-egg nature of the problem. To understand what a user is wanting to do, we need to be able to interpret their actions. But to interpret their actions, we first need to know what the user is trying to do. Hence, during calibration, the user is explicitly asked to produce examples of action-meaning pairs that are used as a reference to learn a tailored action-to-meaning mapping.

A curious mind might ask: "Could we go around this chicken-and-egg problem? Could we allow users to start controlling a machine using their preferred action-to-meaning mapping without knowing it in advance?" An interface capable of solving this problem is called self-calibrating and has been explored in the field of brain-computer interfaces [2]–[7] by relying on the subject's internal consistency.

For this demonstration, we implemented self-calibrating principles to the PIN-entry method from Roth et al. in [1], which makes use of two colored buttons to selectively refine the user password via an elimination process. In [1], those buttons are clearly labelled by color. By introducing self-calibration, we can remove the colors on the buttons and allow the user to choose which color to attribute to each button in their mind.

We call our method IFTT-PIN, which stands for 'If This Then PIN', because it makes extensive use of simple 'If This Then That' reasoning. When buttons are of known colors, such reasoning is as simple as: *"If the user pressed the button B, then they indicated that their digit is of color C, thus their digit is among the set of digits currently colored in C."* The self-calibrating version flips this reasoning on its head and on a per digit case: *"If the user is trying to type the digit D, which is currently colored in C, then when they used the button B, they meant the color C"*, which, when combined with a test of user consistency: *"If, for a particular digit D, the user pressed the same button B to mean two different colors (C1 and C2), then they are not entering the digit D"*, enables us to solve the self-calibration problem.

We first present our implementation of the PIN-Entry method from [1] followed by its self-calibrating version. All our interactive demos are available online with links provided in section 2.3 and 3.3.

## 2 PIN-ENTRY METHOD

IFTT-PIN follows the same principle as the PIN-entry method from Roth et al. [1]. Quoting directly from [1]: *"The principal idea is to present the user the PIN digits as two distinct sets e.g., by randomly coloring half of the keys black and the other half white. The user must enter in which set the digit is by pressing either a separate black or white key. Multiple rounds of this game are played [...]. The verifier e.g., the automatic teller machine (ATM), determines the entered PIN digits by intersecting the chosen sets."*

### 2.1 Interface design

Our interface is split into three parts (Figure 1). The top part displays the PIN. The middle part shows the possible digits (0 to 9) colored in yellow or grey. This is the machine asking: "What color is the digit you want to type?". The bottom part is for the user to answer that question. In [1], the user can click on two colored buttons. Here, the left button is yellow, and the right button is grey. This is the user answering: "My digit is yellow" or "My digit is grey".

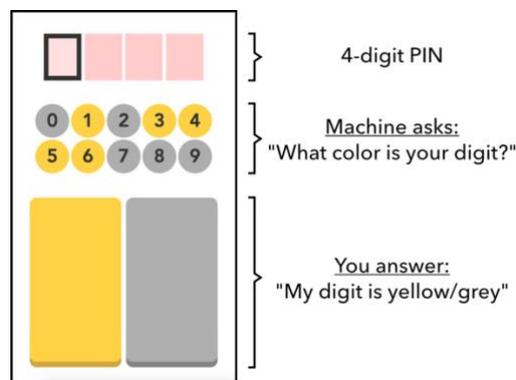

*Figure 1: Breakdown of the interface*

### 2.2 Algorithmic principle

The interface identifies the user's digit as follows: *"If the user presses the left button, then it indicates that their digit is currently yellow, thus their digit is among the yellow-colored digits and all the grey digits can be discarded."* By iteratively changing the color applied to each digit, we can narrow the possible digits down to the one the user has in mind. Interested readers can refer to the pseudocode in [1] (see their Figure 2) for formal notations.

### 2.3 Interactive demonstration

This demo is available at https://jgrizou.github.io/IFTT-PIN/interaction_1.html. To visualize the decision process, we developed a side dashboard that displays the history of your clicks with respect to each digit, see https://jgrizou.github.io/IFTT-PIN/interaction_1_sidepanel.html. A walkthrough video is available at https://youtu.be/6wgOa380uEo.

**3   SELF-CALIBRATION METHOD**

IFTT-PIN adds self-calibration to this interface, and thus removes colors from the buttons. The user decides the color of each button in their mind and uses them as such - never explicitly telling the machine about it. To adequately show the flexibility offered by self-calibration, we increased the number of buttons from 2 to 9 which expands the number of possible color patterns from 2 to 510[1], as illustrated in Figure 2.

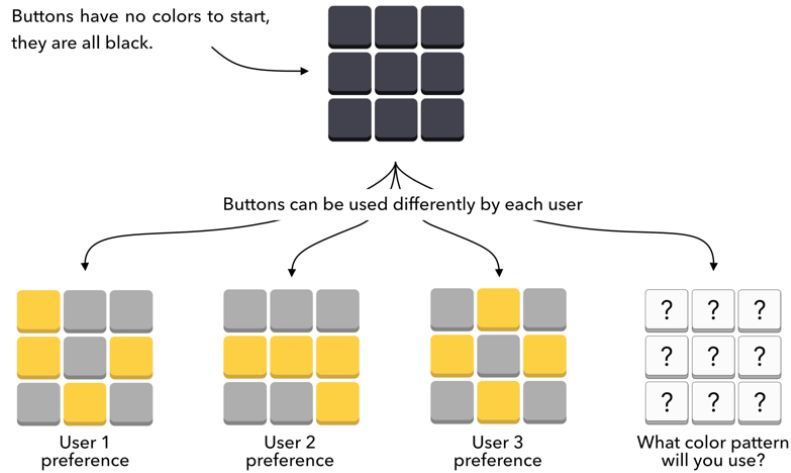

*Figure 2: Example of choice of button-to-color mapping. Each user can assign colors to each button as they prefer and use them as such. At least one button should be assigned for the yellow and grey colors.*

**3.1   Algorithmic principle**

The reasoning used in section 2.2 does not work anymore. However, we know that the user is trying to type one of the ten digits and that one button can only be assigned one color, otherwise the user would be inconsistent. We thus formulate ten different hypotheses of the digit the user is trying to enter, interpret the user's action according to each hypothesis and verify if the user is being consistent. The user can only remain consistent with one hypothesis (the digit they are entering). When only one hypothesis remains consistent, we know that it is the digit the user has in mind and, by extension, the colors assigned to each button.

**3.2   Algorithmic illustration**

This process is illustrated in Figure 3 after one, four, and eight clicks from a typical interaction. For visual clarity, we only show the process for digit 0, 1, 2, and 3, as if the user could only enter one of those four digits.

After one click, the top-left button is marked with a dot because the user just clicked on that button. This dot is yellow for digit 0 and 3, and grey for digit 1 and 2, because, at the time the user clicked on the button, the digits 0 and 3 were yellow, and digits 1 and 2 were grey. After one click, all hypotheses are consistent.

---

[1] With 2 buttons only be yellow-gray or gray-yellow are possible. The number of valid color combination is $2^N - 2$ where N is the number of buttons available. At least one button of each color must be available to the user, so we need to remove the two cases when all buttons are of the same color (all gray or all yellow). With N=9, we have 510 valid color combinations.

After four clicks, the middle button has received two clicks from the user. If the user was entering digit 0 and 2, they would have used the middle button to mean alternatively yellow and grey, which would be a breach of the consistency assumption. Thus, the user is not entering digit 0 or 2. While the middle button would be yellow if the user was typing a 1, and grey if the user was typing a 3, both options remain consistent at this stage.

After eight clicks, the top-left button would have been used for both yellow and grey if the user was entering digit 1. Only digit 3 remains fully consistent, thus we can conclude that the user is trying to enter the digit 3.

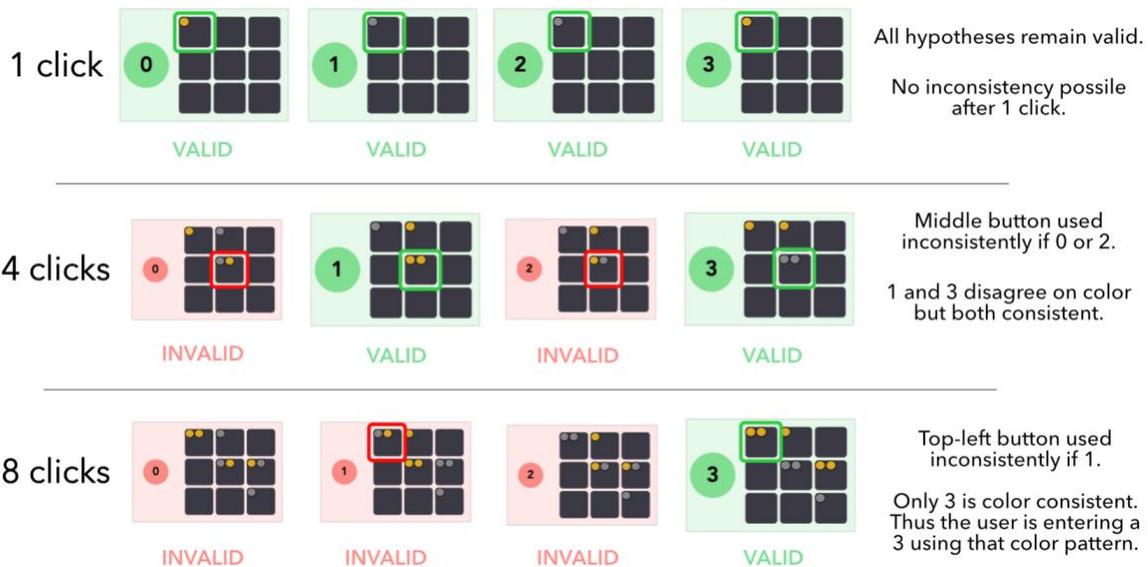

*Figure 3: Illustration of inconsistency detection for digits 0 to 3 after one, four, and eight clicks from a typical interaction.*

### 3.3 Interactive demonstration

IFTT-PIN is available at https://jgrizou.github.io/IFTT-PIN/interaction_2.html. The version with the side dashboard is available at https://jgrizou.github.io/IFTT-PIN/interaction_2_sidepanel.html. A walkthrough video is available at https://youtu.be/t7MQoBnzryQ.

## 4 LIVE INTERACTIVE ONLINE DEMOS

PIN entering is a familiar task and IFTT-PIN can be demonstrated live and interactively in less than 3 minutes. During live demos, we ask members of the audience to choose a PIN and colors to assign to each button. We then enter their chosen PIN using their chosen color pattern. After a few clicks, both their PIN and their colors will appear on screen. Having grasped their attention, we explain how the interface works using the side explanatory panel.

## 5 CRACK THE PIN CHALLENGE

To challenge your understanding of IFTT-PIN, try cracking the PIN entered by a user from a shoulder-surfing video available at https://youtu.be/v8sNuhXb_J4. You can verify you have cracked the correct PIN at https://jgrizou.github.io/IFTT-PIN/challenge.html.

## 6 CONCLUSION

We presented IFTT-PIN, an online interactive PIN-Entry interface conceived to introduce the self-calibration paradigm. IFTT-PIN allows users to enter the PIN of their choice via an elimination process by indicating the color assigned to their digit (yellow or grey). To express their choice, users click on a button whose color is the same as their digit. But buttons do not have any color assigned to them at the start of the interaction. Users are free to define the color of each button in their mind and use them as such without informing the interface. After a few iterations, the interface can infer both the digit the user had in mind and the color of each button used. IFTT-PIN demonstrates a new interactive experience where users can actively decide how to use an interface on-the-fly. Our hope is to inspire the community to invent novel interaction experiences that leverage the self-calibration paradigm.

## ACKNOWLEDGMENTS

We thank the CRI for the fellowship that allowed us to develop this demo. Many thanks to Edwin Paquiot and Joanna May Lee for their design advice, as well as to students and staff at CRI for testing this demo extensively. This work was submitted as a demo to UIST 2021 (https://uist.acm.org/uist2021/) and not accepted. No feedback was provided due to the size of the conference.